\documentclass[a4paper,11pt]{article}

\usepackage{jheppub} 

\usepackage[T1]{fontenc} 

\title{\boldmath On the Study of Non-Relativistic Spin Half-Integer Systems}

\author[a,1]{Hadi Sobhani \note{Corresponding author.}}
\author[a]{Hassan Hassanabadi}

\affiliation[a]{Physics Department, Shahrood University of Technology, Shahrood, Iran.}

\emailAdd{hadisobhani8637@gmail.com}

\abstract{In this article, after introducing appropriate equation for non-relativistic spin half-integer system, Lagrange density of such system has been derived. Then time-evolution of this system in presence of a time-dependent interaction has been done.}

\begin{document} 
\maketitle
\flushbottom

\section{Introduction}
\label{sec:intro}

Schr\"{o}inger equation is the most famous equation in non-relativistic quantum mechanics. With the aim of this equation we are able to investigate many different systems in presence of various situations \cite{1,2}. There is vacant place in non-relativistic quantum mechanics, that is, in this formalism we can’t consider spin of particles in interactions similar what we do in relativistic version of quantum mechanics \cite{3}.  Recently Ajaib published an article in which a fundamental form Schrödinger equation has been derived as \cite{4}

\begin{equation}\label{1}
 - i{\partial _z}\psi  = \left( {i\eta {\partial _t} + {\eta ^\dag }m} \right)\psi.
\end{equation}

Actually the author iterated Eq. (1) to obtain Schr\"{o}dinger equation. It yields to have

\begin{equation}\label{2}
{\left( { - i{\partial _z}} \right)^2}\psi  = {\left( {i\eta {\partial _t}} \right)^2}\psi  + im\left\{ {\eta ,{\eta ^\dag }} \right\}{\partial _t}\psi  + {\left( {{\eta ^\dag }} \right)^2}{m^2}\psi,
\end{equation}

then in order to reach Schrödinger equation, the parameters have to obey below conditions

\begin{equation}\label{3}
\begin{array}{l}
{\eta ^2} = 0,\\\\
{\left( {{\eta ^\dag }} \right)^2} = 0,\\\\
\left\{ {\eta ,{\eta ^\dag }} \right\} = 2{I}.
\end{array}
\end{equation}

In which $I$  is the unit matrix and 

\begin{equation}\label{4}
\eta  = \frac{1}{{\sqrt 2 }}\left( {\begin{array}{*{20}{c}}
	0&{ - i}&0&{ - 1}\\
	{ - i}&0&1&0\\
	0&1&0&{ - i}\\
	{ - 1}&0&{ - i}&0
	\end{array}} \right) = \frac{i}{{\sqrt 2 }}\left( {\begin{array}{*{20}{c}}
	{{\sigma _1}}&{{\sigma _2}}\\
	{ - {\sigma _2}}&{{\sigma _1}}
	\end{array}} \right) = \frac{i}{{\sqrt 2 }}\left( {\begin{array}{*{20}{c}}
	0&{I + {\sigma _2}}\\
	{I - {\sigma _2}}&0
	\end{array}} \right),
\end{equation}

Where ${\sigma _i}(i = 1,2)$  are Pauli matrixes and $\eta $   is non-Hermitian $({\eta ^\dag } = {\eta ^*} \ne \eta )$ and symmetric matrix $({\eta ^T} = \eta )$.

Eq. (1) can be extended in 3 dimensional by writing

\begin{equation}\label{5}
\left( {i({\mu _j}{\partial _j} + \eta {\partial _t}) + {\eta ^\dag }m} \right)\psi  = 0, \qquad (j = 1,2,3)
\end{equation}

in which the parameters are

\begin{equation}\label{6}
\begin{array}{l}
{\mu _1} = i{\gamma _1}{\gamma _2} = \left( {\begin{array}{*{20}{c}}
	{{\sigma _3}}&0\\
	0&{{\sigma _3}}
	\end{array}} \right),\\ \\
	{\mu _2} = {\gamma _0}{\gamma _2} = \left( {\begin{array}{*{20}{c}}
		0&{{\sigma _2}}\\
		{{\sigma _2}}&0
		\end{array}} \right),\\ \\
		{\mu _3} = {\gamma _2}{\gamma _5} = \left( {\begin{array}{*{20}{c}}
			{{\sigma _2}}&0\\
			0&{ - {\sigma _2}}
			\end{array}} \right),
			\end{array}
\end{equation} 

with the famous gamma matrices. In this article, we derive the Lagrange density corresponding Eq. (\ref{1}) in section 2. Then in Sec. 3 we derive  time evolution relation of Eq. (\ref{1}) to probe the time evolution of such system. Section 4 contains theory of Lewis-Riesenfeld dynamical invariant method and details of deriving wave function in presence of time-dependent interaction. In the last section using the Lie algebra method we also derive the time evolution operator.

\section{The Lagrange Density}

In the quantum field theory the start point of any problem or system is a simple concept, the
Lagrange density $\ell ({\psi _\sigma },\frac{{\partial {\psi _\sigma }}}{{\partial {x_\mu }}})$ that Lagrange function can be obtained by integrating over the
three space

\begin{equation}\label{7}
L = \int\limits_V^{} {\ell ({\psi _\sigma },\frac{{\partial {\psi _\sigma }}}{{\partial {x_\mu }}})} {d^3}x,
\end{equation}

where,$\ell ({\psi _\sigma },\frac{{\partial {\psi _\sigma }}}{{\partial {x_\mu }}})$ as is shown in our notation, is function of wave field and all derivatives
because of not lead nonlocal theories, the higher order of derivatives are not considered \cite{3}. Following the variational principle, it is possible to write

\begin{equation}\label{8}
\delta \int {Ldt = } \delta \int {\ell ({\psi _\sigma },\frac{{\partial {\psi _\sigma }}}{{\partial {x_\mu }}}){d^4}x}  = 0,
\end{equation}

which results in

\begin{equation}\label{9}
\frac{\partial }{{\partial {x^\mu }}}\frac{{\partial \ell }}{{\partial (\partial {\psi _\nu }/\partial {x^\mu })}} - \frac{{\partial \ell }}{{\partial {\psi _\nu }}} = 0,
\end{equation}

to have equation of motion. Let us to bring a simpler form of Eq. (\ref{9}) as \cite{3}

\begin{equation}\label{10}
\frac{\partial }{{\partial t}}\frac{{\partial \ell }}{{\partial ({{\dot {\bar{ \psi }}}_\nu })}} + \nabla .\frac{{\partial \ell }}{{\partial (\nabla {{\bar \psi }_\nu })}} - \frac{{\partial \ell }}{{\partial {{\bar \psi }_\nu }}} = 0,
\end{equation}

where dot is time partial derivative. Comparing with Eq. (\ref{1}), we find out that

\begin{equation}\label{11}
i\eta \frac{{\partial \psi }}{{\partial t}} = \frac{\partial }{{\partial t}}(\frac{{\partial \ell }}{{\partial \dot {\bar{\psi }}}}),
\end{equation} 

or

\begin{equation}\label{12}
\ell  = i\eta \dot {\bar {\psi}} \psi  + f(\nabla \bar \psi ,\bar \psi ),
\end{equation}

that $f(\nabla \bar \psi ,\bar \psi )$ should be determined. Using Eq. (\ref{12}) and another comparison Eqs. (\ref{10}) and (\ref{1}) results

\begin{equation}\label{13a}
\nabla .(\frac{{\partial \ell }}{{\partial (\nabla \bar \psi )}}) = i{\mu _j}\nabla \psi ,
\end{equation}

so that 

\begin{equation}\label{13b}
\frac{{\partial f(\nabla \bar \psi ,\bar \psi )}}{{\partial (\nabla \bar \psi )}} = i{\mu _j}\psi ,
\end{equation}

which reads

\begin{equation}\label{14}
f(\nabla \bar \psi ,\bar \psi ) = i{\mu _j}\nabla \bar \psi \psi  + g(\bar \psi ),
\end{equation}

Where $g(\bar{\psi} )$  will be ascertained. Substituting Eq. (\ref{14}) into (\ref{12}), caused to rewrite Eq. (\ref{12}) clearly

\begin{equation}\label{15}
\ell  = i\eta \dot {\bar {\psi}} \psi  + i{\mu _j}\nabla \bar \psi \psi  + g(\bar \psi ).
\end{equation}

For the last time, we again compare Eqs. (\ref{10}) and (\ref{1}) to gain

\begin{equation}\label{16}
 - \frac{{\partial \ell }}{{\partial \bar \psi }} = \frac{{\partial g(\bar \psi )}}{{\partial \bar \psi }} = {\eta ^\dag }m\psi ,
\end{equation} 

that easily results 

\begin{equation}\label{17}
g(\bar \psi ) =  - {\eta ^\dag }m\bar \psi \psi ,
\end{equation}

Finally, inserting Eq. (\ref{17}) into Eq. (\ref{12}) leads to the Lagrange density 

\begin{equation}\label{18}
\ell  = i\eta \dot {\bar {\psi}} \psi  + i{\mu _j}\nabla \bar \psi \psi  - {\eta ^\dag }m\bar \psi \psi.
\end{equation}

\section{Time Evolution Relation}
Since Ajaib showed that Eq. (\ref{1}) can turn into Schrödinger equation, we thought that it has the
ability to have similar time evolution relation as Schrödinger equation does. To obtain such
relation first we should be multiplied Eq. (\ref{1}) by ${\eta ^\dag }$  from the left and using Eq. (\ref{4}), it reaches 

\begin{equation}\label{19}
 - i{\eta ^\dag }{\partial _z}\psi  = i{\eta ^\dag }\eta {\partial _t}\psi ,
\end{equation}

as well as taking complex conjugate of Eq. (\ref{1}) then multiply it by $\eta$ from the left caused to write the final form as

\begin{equation}\label{20}
 - i\eta {\partial _z}\psi  = i\eta {\eta ^\dag }{\partial _t}\psi ,
\end{equation}

Summing Eq. (\ref{19}) and (\ref{20}) and with aim of Eq. (\ref{3}) we obtain

\begin{equation}\label{21}
i{\partial _t}\psi  = \frac{{ - i}}{2}\left( {\eta  + {\eta ^\dag }} \right){\partial _z}\psi,
\end{equation}

As is mentioned in Ref. \cite{4}, the parentheses is equal to $ - i\sqrt 2 {\gamma _2}$  , the  $\gamma$ is the gamma matrices. Rewriting Eq. (\ref{21}) helps us to get to the time evolution relation as

\begin{equation}\label{22}
i{\partial _t}\psi  = {H_{F - Sch}}\psi ,
\end{equation}

where

\begin{equation}\label{23}
{H_{F - Sch}} = \frac{{i{\gamma _2}}}{{\sqrt 2 }}{P_z} + V(z,t).
\end{equation}

So, Eq. (\ref{23}) by the potential as $V(z,t) = f(t)z$  in which $f(t)$  is an arbitrary function of time has the form of

\begin{equation}\label{24}
{H_{F - Sch}} = \frac{{i{\gamma _2}}}{{\sqrt 2 }}{P_z} + f(t)z.
\end{equation}

Now, we are in position to probe time evolution of Eq. (\ref{24}).

\section{Time Evolution and Lewis-Riesenfeld Approach}

In 1969, Lewis and Riesenfeld introduced an approach which is able to investigate time evolution of time dependent systems \cite{5}. In this manner, it is supposed that there exist a Hermitian and invariant operator such as $\hat{I}$ that has time evolution as

\begin{equation}\label{25}
\frac{{d\hat I(t)}}{{dt}} = \frac{{\partial \hat I(t)}}{{\partial t}} + \frac{1}{i}[\hat I(t),H(t)] = 0,
\end{equation}

Acting Eq. (\ref{25}) from the left on the $\left| \psi  \right\rangle $  , after some calculations, it is possible to write

\begin{equation}\label{26}
i\hbar \frac{{\partial (\hat I\left| \Psi  \right\rangle )}}{{\partial t}} = H(\hat I\left| \Psi  \right\rangle ),
\end{equation}

Eq. (26) is showing that the action $\hat{I}(t)$ also satisfies the time evolution equation of the ket.
On the other hand, the ket can be expressed in terms of Eigen kets of $\hat{I}(t)$ as

\begin{equation}\label{27}
\left| \Psi  \right\rangle  = \sum\limits_{\lambda ,\kappa } {{c_{\lambda ,\kappa }}{e^{i{\alpha _{\lambda ,\kappa }}(t)}}} \left| {\lambda ,\kappa ;t} \right\rangle ,
\end{equation}

where $\lambda$ is the eigen value, $\kappa$ denotes the other quantum numbers, $c_{\lambda, \kappa}$ is constant and $\alpha_{\lambda, \kappa}(t)$ derived from the below equation

\begin{equation}\label{28}
\hbar \frac{{d{\alpha _{\lambda ,\kappa }}(t)}}{{dt}} = \left\langle {\lambda ,\kappa } \right|i\hbar \frac{\partial }{{\partial t}} - H(t)\left| {\lambda ,\kappa } \right\rangle.
\end{equation}

By this theory and following the methods in Refs \cite{5,6}, we derived the dynamical invariant for our considered system, Eq. (\ref{24}), as 

\begin{equation}\label{29}
\hat I(t) = {\varepsilon _1}{P_z} + {\varepsilon _2}z + {\varepsilon _3}(t),
\end{equation}

with constants $\varepsilon_1$ and $\varepsilon_2$  and

\begin{equation}\label{30}
{\varepsilon _3}(t) = \frac{{ - i{\gamma _2}}}{{\sqrt 2 }}{\varepsilon _2}t + {\varepsilon _1}\int {f(t)dt + c', \qquad{\rm{(c' = constant)}}}. 
\end{equation}

It is reasonable to consider that because $\hat{I}(t)$  produces a first order differential equation for its own eigen functions, we propose the general form of the eigen function as below

\begin{equation}\label{31}
{\phi _\lambda }(z,t) = \exp \left[{\mu _1}(t)z + {\mu _2}(t){z^2} \right],
\end{equation}

then we propose wave function according the eigen function in form of

\begin{equation}\label{32}
{\psi }(z,t) = K(t)\exp \left[{\mu _1}(t)z + {\mu _2}(t){z^2} \right].
\end{equation}

Using(\ref{32}) and (\ref{24}), it is easy to derived

\begin{equation}\label{33}
\begin{array}{l}
{\mu _1}(t) =  - i \left( \sqrt 2 {\gamma _2}{\mu _2}t + \int {f(t)dt} \right)+ c'',\qquad(c'' = {\rm{constant}}), \\\\
{\mu _2} = {\rm{constant}},\\\\
K(t) = \exp \left[ \frac{{ - i{\gamma _2}}}{{\sqrt 2 }}\int {{\mu _1}(t)dt + c'''} \right],\qquad(c''' = {\rm{constant}}).
\end{array}
\end{equation}

It is instructive that as special cases we consider two cases as

\begin{itemize}
	\item \textbf{Case A:}$f(t) = {d_1}{e^{i\omega t}},\,\,(\omega ,{d_1} = {\rm{constant}})$
	
	This case represents periodic phenomena. So Eq.(\ref{33}) changes to 
	
	\begin{equation}\label{34}
	\begin{array}{l}
	{\mu _1}(t) =  - i \left( \sqrt 2 {\gamma _2}{\mu _2}t + \frac{{{d_1}}}{{i\omega }}{e^{i\omega t}} \right) + {d_2},\,\,\,({d_2} = {\rm{constant}}), \\\\
	
	K(t) = \exp \left[ {\frac{{ - i{\gamma _2}}}{{\sqrt 2 }}\left( { - i\frac{{{\gamma _2}{\mu _2}{t^2}}}{{\sqrt 2 }} - \frac{{{d_1}}}{{{\omega ^2}}}{e^{i\omega t}} + {d_2}t} \right) + {d_3}} \right],\,\,\,({d_3} = {\rm{constant}}).
	\end{array}
	\end{equation}
	
	\item \textbf{Case B:}$f(t) = {g_1}{t^s},\,\,(s,{g_1} = {\rm{constant}})$
	
	This case represents all linear or harmonics interactions. By the way Eq.(\ref{33}) turns into
	
	\begin{equation}\label{35}
	\begin{array}{l}
	{\mu _1}(t) =  - i \left( \sqrt 2 {\gamma _2}{\mu _2}t + \frac{{{g_1}}}{{s + 1}}{t^{s + 1}} \right) + {g_2},\,\,\,({g_2} = {\rm{constant}} ),\\\\
	K(t) = \exp \left[ {\frac{{ - i{\gamma _2}}}{{\sqrt 2 }}\left( { - i\frac{{{\gamma _2}{\mu _2}{t^2}}}{{\sqrt 2 }} - \frac{{{g_1}}}{{(s + 1)(s + 2)}}{t^{s + 2}} + {g_2}t} \right) + {g_3}} \right],\,\,\,({g_3} = {\rm{constant}}).
	\end{array}
	\end{equation} 
\end{itemize}

\section{Time Evolution Operator And  Lie Algebra Method}

Other aspect of time evolution of a system is finding time evolution operator for the system. Having such an operator causes to have wave function in any desired time. According method of Ref. \cite{7} the Ansatz for this operator is as

\begin{equation}\label{36}
U(t) = \exp \left[ {{\xi _1}(t){P_z} + {\xi _2}(t)z + {\xi _3}(t)} \right],
\end{equation}

in which arbitrary functions ${\xi _i}(t),(i = 1,2,3)$  can be derived from

\begin{equation}\label{37}
i{\partial _t}U(t){U^{ - 1}}(t) = H.
\end{equation}

So, we have

\begin{equation}\label{38}
{\partial _t}U(t){U^{ - 1}}(t) = {\dot \xi _1}(t){P_z} + {\dot \xi _2}(t){e^{{\xi _1}(t){P_z}}}\left( z \right){e^{ - {\xi _1}(t){P_z}}} + {\dot \xi _1}(t).
\end{equation}

Following Baker–Hausdorff formula \cite{7}, we can write as 

\begin{equation}\label{39}
\begin{array}{l l}
{e^{{\xi _1}(t){P_z}}}\left( z \right){e^{ - {\xi _1}(t){P_z}}} &= z + {\xi _1}(t)[{P_z},z],\\\\
 &= z - {\xi _1}(t)i.
\end{array}
\end{equation}

Inserting (\ref{39}) into (\ref{38}) and some sorting result

\begin{equation}\label{40}
i\left( {{{\dot \xi }_1}(t){P_z} + {{\dot \xi }_2}(t)z + \frac{{{\xi _1}(t){{\dot \xi }_2}(t)}}{i} + {{\dot \xi }_3}(t)} \right) = \frac{{i{\gamma _2}}}{{\sqrt 2 }}{P_z} + f(t)z,
\end{equation}

equating each coefficient, we have

\begin{equation}\label{41}
\begin{array}{l}
{\xi _1}(t) = \frac{{{\gamma _2}}}{{\sqrt 2 }}t + {q_1},\\\\
{\xi _2}(t) =  - i\int {f(t)dt}  + {q_2},\\\\
{\xi _3}(t) =  - \frac{{{\gamma _2}}}{{\sqrt 2 }}\int {f(t)dt}  + {q_3},
\end{array}
\end{equation}

In which ${q_i},(i = 1,2,3)$  are constant and having the explicit form of $f(t)$ , the exact form of ${\xi _i},(i = 1,2,3)$  can be derived.



\end{document}